\documentclass[11pt]{article}
\usepackage[symbol]{footmisc}

\usepackage[left=2.5cm,top=2.5cm,right=2cm,bottom=2.5cm]{geometry}
\usepackage{amsmath, amssymb}

\usepackage{graphicx,subfigure,epsfig,epsf,color}
\usepackage{slashed}
\usepackage{feynmf}
\usepackage{graphicx}
\usepackage{graphics}
\usepackage{epstopdf}
\usepackage{subfigure}
\usepackage{amsfonts}
\usepackage{sectsty}
\usepackage{hyperref}
\usepackage{cite}
\usepackage{pdflscape}
\usepackage{lipsum}
\begin{document}
 \date{}

\title{Entropies of the various components of the universe}
\maketitle
 \begin{center}
 Hao Yu$~^{a,}$\footnote{yuhaocd@cqu.edu.cn},~Yu-Xiao Liu~$^{b,c,}$\footnote{liuyx@lzu.edu.cn},~Jin Li~$^{a,}$\footnote{cqujinli1983@cqu.edu.cn, corresponding author} \\
 \end{center}
 
 \begin{center}
 $^a$ Physics Department, Chongqing University, Chongqing 401331, China\\
 $^b$ Institute of Theoretical Physics $\&$ Research Center of Gravitation, Lanzhou University, Lanzhou 730000, China\\
 $^c$ Lanzhou Center for Theoretical Physics, Key Laboratory of Theoretical Physics of Gansu Province,
 School of Physical Science and Technology, Lanzhou University, Lanzhou 730000, China
 \end{center}
 \vspace*{0.25in}
 \begin{abstract}
 In this work, we study the entropies of photons, dust (baryonic matter), dark matter, and dark energy in the context of cosmology. When these components expand freely with the universe, we calculate the entropy and specific entropy of each component from the perspective of statistics. Under specific assumptions and conditions, the entropies of these components can satisfy the second law of thermodynamics independently. Our calculations show that the specific entropy of matter cannot be a constant during the expansion of the universe except for photons. When these components interact with the space-time background, there could exist the phenomenon of particle production (annihilation). We study the influence of the interaction on the entropies of these components, and obtain the conditions guaranteeing that the entropy of each component satisfies the second law of thermodynamics.
 \end{abstract}

\section{Introduction}
Since it was found that the universe is expanding, the entropy problem and heat death of the universe have been intriguing topics about the past, present, and future entropies of the universe~\cite{Tolman:1931zz,Tolman2}. The entropy problem includes two aspects involving the past and present entropies of the universe. Although physicists have predicted that entropy production mechanism in the early universe may be associated with vacuum energy, Higgs particles, shear and bulk viscosities of the cosmological fluid~\cite{Weinberg:1971mx,Zeldovich:1972zz,Caderni:1977rd,Morikawa:1984dz}, they have not explained satisfactorily why the initial state of the universe possesses low entropy. This question about the past entropy of the universe is also known as the Boltzmann-Penrose question (see~\cite{Banks:2007ei} and references therein). For the present entropy of the universe, the problem boils down to why the value of the entropy is enormous, which is also closely related to the horizon problem~\cite{Kolb:1990vq}. In 2008, Frampton et al. calculated the total entropy of the supermassive black holes inside the observable universe and found it is the largest contributor to the entropy of the universe~\cite{Frampton:2007xq,Frampton:2008mw,Frampton:2009nx}. Based on the works of Frampton et al., the following year Egan and Lineweaver took into account the latest measurements of the supermassive black hole mass function at the time, and found that the total entropy of the supermassive black holes inside the observable universe is of about $3.1^{+3.0}_{-1.7}\times10^{104}\,k$~\cite{Egan:2009yy}, which is an unimaginably enormous number. As for the cosmic heat death, it presents a gloomy prediction on the future entropy and the fate of the universe~\cite{Thomson:1857,Dyson:1979zz,Adams:1996xe}. The study on the cosmic entropy shows that, the entropy of an expanding universe generally satisfies the second law of thermodynamics~\cite{Morikawa:1985mf,Pavon:2012qn,Mimoso:2013zhp}. Therefore, with the expansion of the universe, all life will eventually disappear, and the universe will be a state of chaos and disorder. In order to avoid the cosmic heat death, people proposed the cyclic model of the universe, in which the law of entropy increase can be inapplicable~\cite{Page:1985ei,Baum:2006ee,Baum:2007de}.

So far, the cosmic entropy has been generalized widely on the basis of classical thermodynamics, which contains horizon entropy~\cite{Gibbons:1977mu, Davies:1987ti}, information entropy~\cite{Hosoya:2004nh,Li:2012qh}, entanglement entropy~\cite{Bak:2012vj,Gonzalez-Diaz:2012ohx,Nakai:2017qos}, and so on. Moreover, there are researches focusing on the entropy of the universe in the presence of particle production (annihilation) and interaction (i.e., energy conversion). In general, particle production necessarily comes with interaction, but interaction does not necessarily lead to the phenomenon of particle production. The common feature of these two kinds of processes is that they can influence the thermodynamic properties of the universe. Therefore, it is significant to study the entropy changes of these processes in cosmology.

The early pioneering works of Parker on particle production provide a microscopic mechanism of particle production in the context of cosmology~\cite{Parker:1968mv,Parker:1969au,Parker:1972kp}, which lays the theoretical foundation for the subsequent study on the entropy of the universe in the presence of particle production. From the perspective of quantum cosmology, particle production could be related to the origin of the universe~\cite{Ford:1978ip,Vilenkin:1982de,Grib:2000mb,Zecca:2012za}, so it may influence the low-entropy state of the early universe. On the issue, the Big Bang theory~\cite{Kawasaki:1994af,Durrer:2000tc,Kawasaki:2004qu,deHaro:2015hdp} and Big Bounce theory~\cite{Novello:2008ra,Quintin:2014oea,Haro:2015zda,
	Brandenberger:2016vhg,Celani:2016cwm} already have multiple mechanisms to create the particles and entropy of the early universe. On the other hand, cosmological particle production could affect the entropy evolution of each component of the universe~\cite{Gibbons:1977mu,Kodama:1981jw,Hu:1986jj,
	Kandrup:1988sg,Calvao:1991wg,Zimdahl:1999tn}. In Refs.~\cite{Prigogine:1986zz,Prigogine:1988zz,Prigogine:1989zz}, the authors proposed to use the second law of thermodynamics to constrain particle production in cosmology. In recent years, the researches on the entropy evolution in cosmology have shown that, using the thermodynamic constraints on the entropy, one can restrain effectively the gravitational theories in which there exists particle production. For example, considering scalar particle production in Horndeski theory~\cite{Horndeski:1974wa}, the coupling coefficients between the scalar field and geometric quantities can be constrained with the second law of thermodynamics and thermodynamic equilibrium~\cite{Yu:2018qzl}. Similarly, if there exists particle production in running vacuum models~\cite{Sola:2013gha,Sola:2014tta}, the change rate of the running vacuum can be constrained by studying the entropy evolution of matter~\cite{SolaPeracaula:2019kfm}.

For most gravitational theories (such as nonminimal coupling theories~\cite{Balakin:2005fu,Harko:2008qz,Harko:2010mv,Harko:2011kv}), if a matter field interacts with other fields (i.e., there is a nonminimal coupling between them), the energy-momentum tensor of the matter field is usually not conserved, unless one redefines the form of the energy-momentum tensor. The nonconservation of the energy-momentum tensor of the matter field, by analogy with the nonconservation of the particle number in an open thermodynamic system, can be explained as irreversible particle production~\cite{Harko:2014pqa,Harko:2015pma,Nunes:2016tsf,Ramos:2017cot,Yu:2018qzl}. In this case, for these gravitational theories, the researches on the entropy evolution of matter in cosmology is similar to the researches on the entropy evolution of cosmological particle production~\cite{Yu:2018qzl,SolaPeracaula:2019kfm}. However, when the matter field interacts with other fields but without particle production, how could we study the entropy evolution of the matter in cosmology? We will discuss the issue in this work.

In addition, by investigating the statistical entropy of matter in a freely expanding universe, we will prove that the specific entropy of matter (except for photons) indeed evolves with the expansion of the universe~\cite{Calvao:1991wg,Ivanov:2019van}, which is directly assumed to be a constant in some literature~\cite{Prigogine:1989zz,Harko:2015pma,Yu:2018qzl}. Taking dust (baryonic matter) and dark matter as the main object of study, we will analyze the difference between the entropy changes in the cases with and without particle production.

The paper is organized as follows. Sec.~\ref{sec2} is devoted to the review of some basic thermodynamic formulas about cosmological particle production (annihilation). In Sec.~\ref{sec3}, we discuss the entropy and specific entropy of each component of the universe when they expand freely. We focus on photons, dust (baryonic matter), dark matter, and dark energy. Then we study the entropies and specific entropies of these components when they have interaction with the space-time background in Sec.~\ref{sec4}. The last part, Sec.~\ref{sec8}, is a summary of our research.

\section{Thermodynamics of cosmological particle production (annihilation)}
\label{sec2}
We consider a homogeneous and isotropic universe and the FLRW metric is given as
\begin{eqnarray}
	\text{d}s^2=-c^2\text{d}t^2+a^2(t)\left(\frac{\text{d}r^2}{1-\tilde k\,r^2}+r^2\text{d}\theta^2+r^2\text{sin}^2\theta \,\text{d}\phi^2\right),
\end{eqnarray}
where $a(t)$ is the scale factor, $c$ is the light speed, and $\tilde k$ represents the curvature of the space. For simplicity, we assume that the universe is spatially flat, i.e., $\tilde k=0$.

Let us review some basic thermodynamic formulas of cosmological particle production (annihilation). Since we focus on the entropy of matter in the universe, it is appropriate to choose the co-moving volume as the thermodynamic system in this work. For a given species of matter, we label the particle number density as $n$, which, combined with the four velocity $U^\alpha$ of the co-moving observer, can be used to define the particle flow as $n^\alpha=n\, U^\alpha$. If there exists a particle production (annihilation) process, then the particle flow satisfies $\nabla_\alpha n^\alpha=\psi$, where $\psi>0$ and $\psi<0$ represent a source and sink of particles, respectively. Since the difference between particle production and annihilation only depends on the sign of $\psi$, here after, they are collectively called particle production for convenience unless otherwise indicated. Then one can define particle production rate as follows:
\begin{eqnarray}
	\Gamma=\frac{\nabla_\alpha n^\alpha}{n}=\frac{\psi}{n}.\label{particlerate}
\end{eqnarray}
For the co-moving observer, the entropy flow vector can be given as
\begin{eqnarray}
	s^\alpha=n\,\sigma\, U^\alpha.\label{entropydefination}
\end{eqnarray}
Note that $\sigma$ is the specific entropy of particles, which denotes the entropy per particle but not per unit mass. To calculate the specific entropy, we review the definition of the Gibbs free energy, which is given by
\begin{eqnarray}
	G(p,T)=U+p\,V-T\,S=\tilde H-T\,S,\label{gibbsfreeenergy}
\end{eqnarray}
where $U$ is the internal energy of the system, $p$ is the pressure, $V$ is the volume, $T$ is the temperature, $S$ is the entropy, and $\tilde H$ is the enthalpy. The integrated Gibbs free energy of the system is the sum of all components:
\begin{eqnarray}
	G_{t}(\tilde p,T)=\sum_{i}U_i+{\tilde p}\,V-T\sum_{i}S_i=\sum_{i}\tilde H_i-T\sum_{i}S_i.\label{gibbsfreeenergy2}
\end{eqnarray}
Here, the evolution of the system is seen as a quasi-equilibrium process. The temperatures of all components are the same and the parameter $\tilde p$ is the total pressure of the system. For such an ideal system, it can be further simplified as follows:
\begin{eqnarray}
	G_{t}(\tilde p,T)=\sum_{i}u_iN_i,\label{gibbsfreeenergy3}
\end{eqnarray}
where we have used $\sum\limits_{i}U_i=T\sum\limits_{i}S_i-\tilde p\,V+\sum\limits_{i}u_iN_i$ (Euler equation). The parameter $u_i$ is the chemical potential of the $i$-th component and $N_i$ is the corresponding particle number. The total internal energy is also equal to $V\sum\limits_{i}\rho_i $, where $\rho_i$ is the energy density of the $i$-th component. Utilizing $\sum\limits_{i}U_i=V\sum\limits_{i}\rho_i $ and Euler equation, the chemical potential of each component satisfies the following relationship:
\begin{eqnarray}
	\sum_{i}u_iN_i=V\sum_{i}\rho_i+\tilde p\,V-T\sum_{i}S_i.\label{chemicalp}
\end{eqnarray}
If there is only one kind of matter in the system, then we can set $i$ to 1. Then the specific entropy of the matter is given by~\cite{Calvao:1991wg}
\begin{eqnarray}
	\sigma =\frac{\rho+p}{T\,n}-\frac{u}{T}.\label{chemicalpone}
\end{eqnarray}
With Eq.~(\ref{chemicalpone}) and Gibbs-Duhem equation ($\sum\limits_{i}S_i\text{d}T-V\,\text{d}\tilde p+\sum\limits_{i}N_i\text{d}u_i=0$), for the system containing only one kind of matter, the differential form of Eq.~(\ref{gibbsfreeenergy2}) can be expressed as
\begin{eqnarray}
	n\,T\, \text{d}\sigma=\text{d}\rho-\frac{\rho+p}{n} \text{d} n.\label{gibbs}
\end{eqnarray}
With this equation and the definition of the entropy flow~(\ref{entropydefination}), the authors in Ref.~\cite{Calvao:1991wg}, for the first time, obtained the following formula in thermodynamics of cosmological particle production:
\begin{eqnarray}
	\nabla_\alpha s^\alpha=\nabla_\alpha (n\,\sigma\, U^\alpha)=\psi\, \sigma +n\, \dot{\sigma}=\frac{\theta}{T}\left(p+\rho+\frac{\dot \rho}{\theta}\right)-\frac{u\, \psi}{T},\label{gibbs2}
\end{eqnarray}
where $\theta=\nabla_\alpha U^\alpha$. For the FLRW metric, $\theta=3H$, where $H$ is the Hubble constant.

Note that the above pressure $p$ contains only the pressure of the fluid itself. In cosmology, however, we usually treat the extraneous term caused by particle production in the equation of motion as an extra pressure, which is called production pressure. We label the production pressure as $p_c$ and then the equation corresponding to the divergence of the energy-momentum tensor of the (perfect) fluid can be written as
\begin{eqnarray}
	\dot\rho+3H(p+\rho+p_c)=0.\label{conservation11}
\end{eqnarray}
Suppose the production pressure satisfies $p_c=-\beta\, \psi/3H$, where $\beta$ is a positive parameter. For an expanding universe with a particle production (not annihilation) process, we have $3H>0$ and $\psi>0$. Then the production pressure should be $p_c<0$, which is in line with the conclusion of most literature~\cite{Ford:1986sy,Traschen:1990sw,Abramo:1996ip,Prigogine:1989zz,Zimdahl:1999tn,Nunes:2015rea,
	Nunes:2016aup,Pan:2016bug}. Taking Eq.~(\ref{conservation11}) into Eq.~(\ref{gibbs2}), we have
\begin{eqnarray}
	\nabla_\alpha s^\alpha&=&\frac{\psi}{T}(\beta-u)\nonumber\\
	&=&\psi\,\sigma+(\beta-\frac{\rho+p}{n})\frac{\psi}{T},
	\label{energyconservation4}
\end{eqnarray}
where we have used Eq.~(\ref{chemicalpone}) to get the second equality. For $\psi>0$ and $\beta>u$, the entropy of particles is growing. If $\beta<u$, we require $\psi<0$ (because of the second law of thermodynamics), which means particles can be only annihilated. It is worth mentioning that $\beta$ and $u$ in fact are not independent. Therefore, the condition $\beta<u$ is not easy to implement, which relies on the specific model of the universe and the species of matter. Contrasting Eq.~(\ref{energyconservation4}) with Eq.~(\ref{gibbs2}), one can get
\begin{eqnarray}
	\dot{\sigma}=\frac{\psi}{n\,T}(\beta-\frac{\rho+p}{n}).\label{sigmadot}
\end{eqnarray}
If the specific entropy $\sigma$ is a constant, we need $\beta=(\rho+p)/n$ or $\psi=0$. $\psi=0$ means the particle number is conserved and the system has always been in a state of thermodynamic equilibrium. If $\beta=(\rho+p)/n$, the production pressure is given as
\begin{eqnarray}
	p_c=-\frac{\rho+p}{3H\,n}\psi.\label{productionpressing}
\end{eqnarray}
According to Eq.~(\ref{gibbs}), $\dot\sigma=0$ yields
\begin{eqnarray}
	\dot\rho=\frac{\dot n}{n}(\rho+p).\label{rhodot}
\end{eqnarray}
With Eqs.~(\ref{conservation11}), (\ref{productionpressing}) and (\ref{rhodot}), we finally obtain $\frac{\dot n}{n}=-3H+\Gamma$, where $\Gamma$ is the particle production rate defined by Eq.~(\ref{particlerate}). Note that here it seems that $\frac{\dot n}{n}=-3H+\Gamma$ is based on $\sigma$ being a constant, but it is easy to prove that the result also holds true for universal $\sigma$ with Eqs.~(\ref{gibbs}), (\ref{conservation11}) and (\ref{sigmadot}). This result actually could be obtained directly from the definition of the particle number inside the co-moving volume, which is given as $N\sim n\, a^3$. Therefore, the intuitive definition of the particle production rate is
\begin{eqnarray}
	\Gamma=\frac{\dot N}{N}=\frac{\dot n\, a^3+3n\,\dot a\, a^2}{n\,a^3}
	=\frac{\dot n }{n}+3H.\label{rateofchange}
\end{eqnarray}

Based on the definition of the specific entropy, the entropy of a given species of matter can be denoted as $S=\sigma\, n\, V=\sigma\, N$. Therefore, the growth rate of the entropy in thermodynamics of cosmological particle production is give by
\begin{eqnarray}
	\frac{\dot S}{S}=\frac{\dot N}{N}+\frac{\dot\sigma}{\sigma}
	=\Gamma+\frac{\dot\sigma}{\sigma}=\frac{\dot n }{n}+3H+\frac{\dot\sigma}{\sigma}.\label{dotentropy1}
\end{eqnarray}
It is convenient to rewrite it in the following form
\begin{eqnarray}
	\frac{\text{d}S}{\text{d}t}=S\left(\frac{\dot\sigma}{\sigma}
	+\Gamma\right).\label{dotentropy2}
\end{eqnarray}
If $\dot\sigma\neq0$, the term $\frac{\dot\sigma}{\sigma}$ will complicate the entropy evolution. For example, it can be seen that even though $\Gamma<0$, $\frac{\text{d}S}{\text{d}t}>0$ can be still true with some special choices of the parameter $\sigma$~\cite{Calvao:1991wg}, which means that particle annihilation could happen under the constraint of the second law of thermodynamics.

\section{The (specific) entropy of matter in the absence of interaction}
\label{sec3}

It is known that for an expanding system in classical thermodynamics, the entropy evolution depends on the way the system expands and the properties of internal matter. Since the expansion of the universe is an intrinsic property of space-time, the boundary of the co-moving volume does not work on its surrounding. Therefore, it is reasonable to suppose that the expansion of the universe is similar to the free expansion of a classical thermodynamic system. For the matter with unknown properties, we can assume that its specific entropy is a constant, so that the entropy of the system is only related to the particle number~\cite{Prigogine:1989zz,Harko:2015pma,Yu:2018qzl}. However, such an assumption is obviously rough, and the specific entropy evolving over time is more in line with the real situation of the universe~\cite{Calvao:1991wg,Ivanov:2019van}. In this section, we assume that there is no interaction between all components of the universe, and we study the entropy and specific entropy of matter in the context of general relativity (that is, the coupling between matter and gravity is a minimal coupling). In this case, we can study different substances individually. Note that the substances we study here are the particles inside the co-moving volume. It is currently unclear whether the boundary of the co-moving volume has an area entropy similar to the horizon entropy. Even if it has an area entropy, we also do not know whether the definition of the area entropy is consistent with the horizon entropy. Therefore, in this work we only consider the particle entropy inside the co-moving volume and ignore the possible area entropy of the co-moving volume.

\subsection{Photons}
\label{sec3.1}

We first consider the cosmic microwave background (CMB) radiation, which could be regarded as black-body radiation. Therefore, the entropy of photons inside the co-moving volume ($V\sim a^3$) is given by~\cite{Baierlein,Herrmann,Leff}
\begin{eqnarray}
	S=\frac{4\pi^2 k^4}{45c^3\hbar^3}V\,T^3=\text{Const.},\label{1545}
\end{eqnarray}
where $k$ is the Boltzmann constant and $T\sim a^{-1}$ is the temperature of photons. Since the entropy of photons is a constant, we can think of photons as being in special thermodynamic equilibrium, which is in accordance with the second law of thermodynamics. The particle number of photons is given as
\begin{eqnarray}
	N=\frac{2 k^3 \zeta(3)}{\pi^2 c^3\hbar^3}V\,T^3=\text{Const.},\label{15451}
\end{eqnarray}
where $\zeta(n)$ is the Riemann zeta function. Therefore, the specific entropy of photons is
\begin{eqnarray}
	\sigma(a)=\frac{S}{N}=\frac{2 \pi^4 k}{45\zeta(3)},\label{15452}
\end{eqnarray}
which is also a constant. The result that the specific entropy of photons is a constant can also be calculated by Eq.~(\ref{chemicalpone}), because the chemical potential of photons is zero~\cite{Baierlein,Herrmann,Leff} and the pressure of photons satisfies $p=\frac{1}{3}\rho$. According to Eq.~(\ref{chemicalpone}), the specific entropy of photons can be given as
\begin{eqnarray}
	\sigma(a) =\frac{\rho+p}{T\,n}=\frac{4}{3}\frac{U}{N\,T}=\frac{2 \pi^4 k}{45\zeta(3)},\label{chemicalpone1122}
\end{eqnarray}
where $U=\frac{\pi^2 k^4}{15c^3\hbar^3}V\, T^4$ is the internal energy of photons. For the matter whose chemical potential and equation of state are unclear, it is not feasible to calculate its specific entropy by Eq.~(\ref{chemicalpone}).

\subsection{Dust (baryonic matter) and dark matter}
\label{sec3.2}

Next, we consider dust (baryonic matter) and dark matter. We still do not figure out all the properties of dark matter. If the particle number density of dark matter in the universe is small enough, then we can treat the dark matter particles as nearly independent particles and ignore their quantum effects. Since such an assumption is not sheer speculation~\cite{Bertone:2004pz}, we can regard dust and dark matter as a class of substances with similar thermodynamic and statistical properties. We take dust as an example to discuss the entropy and specific entropy of this class of substances. Since the matter with vanishing pressure has no thermodynamic effect, to study the entropy of dust, we treat it as a perfect fluid with a weak pressure. In addition, if the volume of each dust particle and the interaction between dust particles can be ignored (i.e., the collisions between dust particles are elastic), it is appropriate to consider dust as a perfect gas. In this case, the equation of state of dust is no longer $p=0$ but
\begin{eqnarray}
	p=\frac{\tilde{n}\, \tilde{R}\, T}{V},\label{equationofstate}
\end{eqnarray}
where $p$ and $T$ are the pressure and temperature of dust, respectively. $V\sim a^3$ is still the co-moving volume. The parameter $\tilde{n}=N/N_A$ is the amount of substance of dust, where $N$ is the total number of dust particles inside the co-moving volume and $N_A=6.022\times10^{23}$ is the Avogadro constant. $\tilde{R}$ is the universal gas constant, whose value is the product of the Boltzmann constant $k$ and Avogadro constant $N_A$. It is convenient to rewrite Eq.~(\ref{equationofstate}) as
\begin{eqnarray}
	p\, V=N\, k\, T.\label{equationofstate2}
\end{eqnarray}
If dust expands freely from the co-moving volume $V_a$ to $V_b$, according to the macroscopic definition of entropy or letting the process be equivalent to a reversible isothermal expansion, the entropy change of dust is given by
\begin{eqnarray}
	\Delta S=S_b-S_a=\int_a^b\frac{\text{d}\hat Q}{T}=\int_a^b\frac{p\,\text{d}V}{T}
	=N\, k(\ln V_b-\ln V_a),\label{entropychange}
\end{eqnarray}
where $\text{d}\hat Q=\text{d}U+p\,\text{d}V$. Since dust expands freely, $\text{d}U=0$ and $T$ is a constant throughout the process. The above formula indicates that the entropy of dust satisfies the second law of thermodynamics in an expanding universe.

From the definition $S=\sigma\, n\, V=\sigma\, N$, $\Delta S$ is also derived as
\begin{eqnarray}
	\Delta S=\sigma_bN-\sigma_aN,\label{entropychange2}
\end{eqnarray}
where $\sigma_b$ and $\sigma_a$ are the specific entropies of dust particles at the beginning and end of the process, respectively. Comparing Eqs.~(\ref{entropychange}) and~(\ref{entropychange2}), we can obtain
\begin{eqnarray}
	\sigma_b-\sigma_a=k(\ln V_b-\ln V_a).\label{entropychange3}
\end{eqnarray}
Due to $V_b\neq V_a$ one obtains $\sigma$ is not a constant. According to Eq.~(\ref{entropychange3}), the specific entropy of dust can be defined as
\begin{eqnarray}
	\sigma(a)=k\,\ln V+\sigma_0=k\,\ln \left(a^3\right)+\sigma_0,
\end{eqnarray}
whose evolution is only related to the scale factor. The parameter $\sigma_0$ depends on the temperature, particle number, and the mass of a single dust particle. Here, dust is regarded as a perfect gas, and so the calculation on its entropy is more concise based on classical thermodynamics. If the entropy of dust is calculated based on statistics, the specific expression of $\sigma_0$ can be determined\footnote{According to classical statistical mechanics, for a perfect gas, the number of microscopic states satisfies $\Omega(N,E,V)\propto V^N$. The entropy of the system is given as $S=k \ln\Omega(N,E,V)$. To calculate $\sigma_0$, we need to figure out the specific form of $\Omega(N,E,V)$. For example, if the perfect gas consists of monatomic molecules, $S=k \ln\Omega(N,E,V)=N\, k\ln[\frac{V}{N}(\frac{2\pi k\,m\,T}{h_0^2})^{3/2}]+\frac{5}{2}N\, k$, where $m$ is the mass of a molecule. Note that $h_0$ is an undetermined constant, because the concept of absolute entropy does not exist in classical statistics. Then $\sigma_0$ is given as $\sigma_0=k\ln[\frac{1}{N}(\frac{2\pi k\,m\,T}{h_0^2})^{3/2}]+\frac{5}{2} k$, where $N$, $m$, and $T$ are all constants for the perfect gas that expands freely.}.

If there exists energy conversion between dust and the space-time background, then the expansion of the universe is not an isothermal process for dust. To calculate the entropy change of dust, one needs to figure out the initial and final states of the system. Note that no matter how the universe expands, the number of dust particles inside the co-moving volume is always conserved in the absence of particle production. If we require $\dot\sigma=0$ (which means that $\Delta S=0$), then the expansion of the universe obeys $\text{d}\hat Q=\text{d}U+p\,\text{d}V=0$. Therefore, in the absence of particle production, the specific entropy of dust can remain constant only when the expansion of the universe is a reversible adiabatic expansion. Since we have assumed that dark matter has similar thermodynamic properties to dust, the above conclusions also apply to dark matter.

\subsection{Dark energy}
\label{sec3.3}

In this section, we discuss the entropy and specific entropy of dark energy. About the thermodynamic properties of dark energy, it has been determined that its pressure is negative, which makes it difficult to calculate its entropy with classical thermodynamics. In this work, we regard it as a substance which obeys universal statistical laws. Therefore, we can calculate the entropy of dark energy with classical statistical mechanics. For dark energy, the particle number density may change with temperature like photons. Without loss of generality, we assume that the number of free dark energy particles evolves with the expansion of the universe, and the corresponding particle production rate is $\Gamma$. If dark energy particles expand freely from the co-moving volume $V_a$ to $V_b$, the particle number is expressed as
\begin{eqnarray}
	N_b=N_a\exp\left[\int_a^b\Gamma \text{d}t\right].\label{newparticle}
\end{eqnarray}

\begin{figure}
	\centering
	\includegraphics[width=11.5cm,height=8cm]{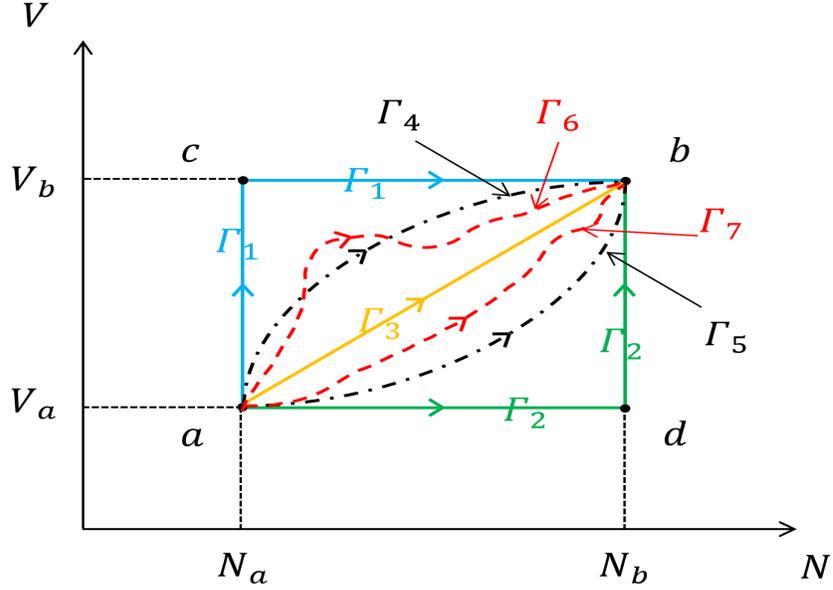}\\
	\caption{Plot of the relationship between the particle production rate ($\Gamma$) of dark energy and the volume of the system. For any $\Gamma_i$, we have $N_b-N_a=\int_a^b\Gamma_i\,\text{d}t$. Since the states of the system are definite at the points $a$ $(V_a, N_a)$ and $b$ $(V_b, N_b)$, for all curves, the entropy changes of the system between the points $a$ and $b$ are the same. The blue solid line ($\Gamma_1$) indicates that the system first expands from $V_a$ to $V_b$ with a conserved particle number, and then the particle number increases from $N_a$ to $N_b$ in the constant volume $V_b$. The processes of the green solid line ($\Gamma_2$) are exactly the opposite of $\Gamma_1$. The solid yellow line ($\Gamma_3$) indicates that the increase of the particle number is proportional to the increase of the volume. The two black dash-dotted lines present that the particle production rate changes with the volume slowly first and fast afterwards ($\Gamma_4$), and fast first and slowly afterwards ($\Gamma_5$), respectively. The two red dashed lines ($\Gamma_6$ and $\Gamma_7$) show the situation where the relationship between the particle production rate and the volume is complicated.}\label{p11}
\end{figure}

Since entropy is a quantity determined by the state of the system, to calculate the entropy change of dark energy in the process, we only need to find out the initial and final states of the system. For the given $N_b$ and $N_a$, there could be different $\Gamma$'s (see Fig.~\ref{p11}). For all lines, the thermodynamic arguments of dark energy at the initial (final) state are the same, so we can calculate the entropy change of dark energy with any of these lines. We choose the blue solid line in Fig.~\ref{p11} to calculate the entropy change of dark energy from the co-moving volume $V_a$ to $V_b$. From point $a$ to point $c$, the particle number inside the co-moving volume does not change. For convenience, we can assume that the statistical properties of dark energy particles are similar to the ones of the perfect monatomic gas. In fact, dark energy is probably nothing but a cosmological effect caused by particle generation, because the process of particle generation will contribute a negative pressure to the Friedmann equations, which could be a possible explanation for the accelerating expansion of the universe~\cite{Ford:1986sy,Traschen:1990sw,Abramo:1996ip,Prigogine:1989zz,Zimdahl:1999tn,Nunes:2015rea,Nunes:2016aup,Pan:2016bug}. Since the particle number of the dark energy we study is non-conserved, even if the dark energy particles are ordinary particles, it could still result in the accelerating expansion of the universe. We will not delve into the details of the issue here, but try to provide some conjecture basis for the hypothesis that it is possible to regard dark matter (which could be the presently known particles with a low particle number density and a non-conserved particle number in the context of cosmology) as a special perfect monatomic gas.

Assuming that the number of microscopic states of dark energy particles with energy between $E$ and $E+\Delta E$ is given as
\begin{eqnarray}
	\Omega(E)= \frac{3N}{2}\frac{\Delta E}{E}\left(\frac{V}{h_0^3}\right)\frac{(2\pi m\, E)^{3N/2}}{N!\left(\frac{3N}{2}\right)!}.
\end{eqnarray}
Here, $h_0$ is an undetermined constant, $m$ is the mass of a single dark energy particle, and $N$ is the number of dark energy particles. According to the definition of Boltzmann entropy ($S=k\ln \Omega$), the entropy of dark energy can be given as
\begin{eqnarray}\label{2222}
	S= N\,k\ln\left[\frac{V}{N}\left(\frac{2\pi k\,m\, T}{h_0^2}\right)^{3/2}\right]+\frac{5}{2}N\,k.
\end{eqnarray}
Therefore, the entropy increase of dark energy in the process from point $a$ to point $c$ is
\begin{eqnarray}\label{aaaaa56646}
	\Delta S_{a\rightarrow c}=N_a k\left\{\ln \left[V_b(m_cT_c)^{3/2}\right]-\ln \left[V_a(m_aT_a)^{3/2}\right]\right\}.
\end{eqnarray}
If energy conservation holds for dark energy\footnote{It is generally believed that only photons and ultra-relativistic particles have the characteristic of energy nonconservation in the expanding universe.}, then the temperature of dark energy particles will remain unchanged. And since the particle number is conserved in the process, we have $m_cT_c=m_aT_a$, which leads to Eq.~\eqref{aaaaa56646} being equal to Eq.~\eqref{entropychange}.

Then from point $c$ to point $b$, it is an isochoric process and the particle number increases from $N_a$ to $N_b$. According to Eq.~\eqref{2222}, the entropy increase of dark energy in the process can be given by
\begin{eqnarray}
	\Delta S_{c\rightarrow b}&=&N_a\exp\left[\int_a^b\Gamma \text{d}t\right] k\ln\left[\frac{V_b}{N_a\exp\left[\int_a^b\Gamma \text{d}t\right]}\left(\frac{2\pi k\, m_b T_b}{h_0^2}\right)^{3/2}\right]+\frac{5}{2}N_a\exp\left[\int_a^b\Gamma \text{d}t\right] k\nonumber\\
	&&-\left(N_a k\ln\left[\frac{V_b}{N_a}\left(\frac{2\pi k\, m_c T_c}{h_0^2}\right)^{3/2}\right]+\frac{5}{2}N_a k\right).
\end{eqnarray}
Then the entropy increase of dark energy from point $a$ to point $b$ is
\begin{eqnarray}\label{316aa}
	\Delta S_{a\rightarrow b}&=&\Delta S_{a\rightarrow c}+\Delta S_{c\rightarrow b}\nonumber\\
	&=&N_a\exp\left[\int_a^b\Gamma \text{d}t\right] k\ln\left[\frac{V_b}{N_a\exp\left[\int_a^b\Gamma \text{d}t\right]}\left(\frac{2\pi k\, m_b T_b}{h_0^2}\right)^{3/2}\right]+\frac{5}{2}N_a\exp\left[\int_a^b\Gamma \text{d}t\right] k\nonumber\\
	&&-\left(N_a k\ln\left[\frac{V_a}{N_a}\left(\frac{2\pi k\, m_a T_a}{h_0^2}\right)^{3/2}\right]+\frac{5}{2}N_a k\right).
\end{eqnarray}
There are many factors that affect the value of $\Delta S_{a\rightarrow b}$, so it is not easy to determine the conditions guaranteeing $\Delta S_{a\rightarrow b}>0$. But if we have
\begin{eqnarray}\label{465456465}
	\exp\left[\int_a^b\Gamma \text{d}t\right]>1\,\,\,\,\,\, \text{and}\,\,\,\,\,\, \frac{V_b}{N_a\exp\left[\int_a^b\Gamma \text{d}t\right]}\left(\frac{2\pi k\, m_b T_b}{h_0^2}\right)^{3/2}>\frac{V_a}{N_a}\left(\frac{2\pi k\, m_a T_a}{h_0^2}\right)^{3/2},
\end{eqnarray}
$\Delta S_{a\rightarrow b}>0$ is tenable and so the second law of thermodynamics is satisfied. On account of energy conservation, $\exp\left[\int_a^b\Gamma \text{d}t\right]>1$ means that $m_b T_b<m_a T_a$. With Eq.~\eqref{entropychange2}, the specific entropy of dark energy particles can be defined as
\begin{eqnarray}\label{h91}
	\sigma(V,N,m,T)= k\ln\left[\frac{V}{N}\left(\frac{2\pi k\, m\, T}{h_0^2}\right)^{3/2}\right]+\frac{5}{2} k,
\end{eqnarray}
which evolves with four parameters of the system composed of dark energy particles. If we can figure out the particle production rate of dark energy during the expansion of the universe and the evolution of other variables over the scale factor, then the specific entropy of dark energy particles will only evolve with the scale factor:
\begin{eqnarray}
	\sigma(a)= k\ln\left[\frac{\frac{4\pi}{3}a^3}{N_0\exp\left[\int_0^a\Gamma(a) \text{d}t\right]}\left(\frac{2\pi k \, m(a) T(a)} {h_0^2}\right)^{3/2}\right]+\frac{5}{2} k,
\end{eqnarray}
where $N_0$ is the number of dark energy particles at the initial state.

Based on the above calculations, it is found that for a freely expanding component of the universe, the specific entropy generally evolves over time except photons. For the co-moving volume, the entropy of photons (which are considered as black-body radiation) is a constant (see Eq.~(\ref{1545})). Regarding dust (or dark matter) as a classical perfect gas, its entropy inside the co-moving volume increases with the expansion of the universe (see Eq.~(\ref{entropychange})). As for dark energy (whose statistical law is assumed to be the same as a classical perfect monatomic gas but with a nonconserved particle number), the entropy inside the co-moving volume could satisfy the second law of thermodynamics under certain constraints (see Eq.~(\ref{465456465})). In the next section, we discuss the entropy changes of these components in the presence of interaction.

\section{The (specific) entropy of matter in the presence of interaction}
\label{sec4}

When matter interacts with an unknown substance (or the space-time background), there must be energy conversion between them. If we do not delve into the microscopic process of the interaction, then the energy conversion may result in two extreme macroscopic phenomena for the matter: A) the particle number is conserved but the energy of a single particle changes; B) the particle number is nonconserved but the energy of a single particle remains unchanged. Both of these situations will lead to a change in the entropy of the matter. In this section, we will analyze how these two different processes affect the entropy of the matter. For convenience, we ignore the entropy of the unknown substance (or the space-time background), which is the common practice employed in similar researches, such as particle production induced by a nonminimal coupling between matter and geometry~\cite{Harko:2014pqa,Harko:2015pma,Nunes:2016tsf,Ramos:2017cot,Yu:2018qzl} or particle production induced by running vacuum~\cite{SolaPeracaula:2019kfm}.

\subsection{Photons}
\label{sec5}

In order to ignore reasonably the entropy of the substance coupled to the CMB radiation, we can assume that the CMB radiation is nonminimally coupled to the space-time background~\cite{Lima:2021rqb}. The thermodynamic state functions for a black-body photon gas indicates that one can describe completely photons inside the co-moving volume with an independent argument: temperature. Here we need to emphasize two properties of photons. First, the particle number of photons must be nonconserved when they are coupled to the space-time background. For photons inside the co-moving volume, the particle number is proportional to the cube of temperature. In the standard $\Lambda$CDM, the temperature of photons is inversely proportional to the scale factor, so the particle number of photons inside the co-moving volume is conserved. If photons are coupled to the space-time background, then the temperature is no longer inversely proportional to the scale factor, which means that the particle number is no longer conserved. Since the particle number of photons inside the co-moving volume is only the function of temperature, when photons are coupled to the space-time background, the two extreme cases mentioned above are completely equivalent on the macroscopic level. Second, the energy of photons is not conserved in the universe. Even if photons are not coupled to any matter (such as the standard $\Lambda$CDM), the energy ($U=\frac{\pi^2 k^4}{15c^3\hbar^3 }V\, T^4$) of photons inside the co-moving volume always decreases with the expansion of the universe.

We suppose that the temperature of the CMB radiation is $T_0$ and the co-moving volume is $V_0$ at some point. After a period of time, the co-moving volume of the universe becomes $V_1$. The temperature of photons becomes $T_1$ due to the expansion of the universe and the interaction. The total energy transformed from the space-time background to photons is labeled as $\bar E$.

\subsection{Dust (baryonic matter) and dark matter}
\label{sec6}

In this section, we still treat dust (baryonic matter) and dark matter as a class of substances which satisfy the statistical law of a perfect monatomic gas, and take dust as an example to study the corresponding entropy and specific entropy in the presence of interaction. Similarly, we assume that dust is nonminimally coupled to the space-time background. Since the universe is homogeneous and isotropic, the interaction can be regarded as occurring everywhere with equal probability in the universe. Then we can suppose that dust is always in quasi-equilibrium. In this case, both of the extremes (with and without particle production) could happen from a macroscopic perspective.

\subsubsection{In the presence of particle production}
\label{sec61}

We first study the entropy change of dust in the presence of particle production. At the initial state, we assume that the rest mass of a dust particle is $m_0$, the temperature of dust is $T_0$, and the co-moving volume is $V_0$. The entropy of dust can be given as
\begin{eqnarray}\label{22223www}
	S_0= N_0k\ln\left[\frac{V_0}{N_0}\left(\frac{2\pi k\, m_0 T_0}{h_0^2}\right)^{3/2}\right]+\frac{5}{2}N_0k,
\end{eqnarray}
where $N_0$ is the particle number. If the total energy transformed from the space-time background to dust is $\bar E$ when the co-moving volume expands from $V_0$ to $V_1$, according to special relativity, the change in the number of dust particles is
\begin{eqnarray}\label{22223www11}
	\Delta N=\frac{\bar E}{m_0\,c^2}.
\end{eqnarray}
Note that the temperature of dust and the rest mass of a dust particle have been assumed to remain constant in the presence of particle production. Therefore, when the co-moving volume is equal to $V_1$, the entropy of dust is given as
\begin{eqnarray}\label{22223www22}
	S_1= (N_0+\Delta N)k\ln\left[\frac{V_1}{N_0+\Delta N}\left(\frac{2\pi k\, m_0 T_0}{h_0^2}\right)^{3/2}\right]+\frac{5}{2}(N_0+\Delta N)k.
\end{eqnarray}
Then the entropy change of dust is equal to
\begin{eqnarray}\label{22223www223}
	S_1-S_0= \Delta N\, k\ln\left[\frac{V_1}{N_0+\Delta N}\left(\frac{2\pi k\, m_0 T_0}{h_0^2}\right)^{3/2}\right]+ N_0 k\ln\left[\frac{V_1}{N_0+\Delta N}\frac{N_0}{V_0}\right]+\frac{5}{2}\Delta N\, k.
\end{eqnarray}

When $\bar E<0$ (the energy of dust is absorbed by the space-time background), the total number of dust particles inside the co-moving volume will decrease, i.e., $\Delta N<0$. Let us check whether the specific entropy of dust will decrease as well. From Eqs.~(\ref{22223www}) and (\ref{22223www22}), it can be found that the difference in specific entropy between the initial and final states satisfies
\begin{eqnarray}\label{22223www11111}
	\sigma_1-\sigma_0= k\ln\left[\frac{V_1}{N_0+\Delta N}\frac{N_0}{V_0}\right]=k\ln\left[\frac{n_0}{n_1}\right],
\end{eqnarray}
where $n_0$ and $n_1$ are the particle number densities of dust at the initial and final states, respectively. Since the co-moving volume is increasing and the particle number is decreasing, we have $\frac{n_0}{n_1}>1$, which means $\sigma_1>\sigma_0$. Therefore, the specific entropy of dust increases as the particle number decreases. Given that, to determine the entropy change of dust, we need to discuss further the parameters in Eq.~(\ref{22223www223}). Note that there are five independent parameters ($V_1$, $V_0$, $N_0$, $\Delta N$, and $m_0T_0$) that could influence the sign of Eq.~(\ref{22223www223}). Taking no account of the constraints of observations, the values of the parameters at least need to satisfy the following fundamental constraints:

I) $S_0>0$ (entropy is positive);

II) $S_1>0$ (entropy is positive);

III) $V_1>V_0>0$ (the universe is expanding);

IV) $-N_0<\Delta N<0$ (the particle number at the final state cannot be negative);

V) $m_0T_0>0$ (the temperature of any matter cannot be absolute zero).

Despite the above constraints, it is still difficult to determine whether Eq.~(\ref{22223www223}) satisfies the second law of thermodynamics. To determine the entropy evolution of dust, we can calculate the time derivative of the entropy of dust. According to Eq.~(\ref{22223www}), it is given as
\begin{eqnarray}\label{211123www}
	\frac{\text{d}S}{\text{d}t}&=& N'k\ln\left[\frac{V}{N}\left(\frac{2\pi k\, m_0 T_0}{h_0^2}\right)^{3/2}\right]
	+N\,k\left(\frac{V'}{V}-\frac{N'}{N}\right)
	+\frac{5}{2}N'k\nonumber\\
	&=&N\,k\left\{\frac{N'}{N}\ln\left[\frac{V}{N}\left(\frac{2\pi k\, m_0 T_0}{h_0^2}\right)^{3/2}\right]
	+\frac{3}{2}\frac{N'}{N}
	+\frac{V'}{V}\right\},
\end{eqnarray}
where the prime indicates the derivative with respect to time. When $N$ tends to 0, dust particles are completely annihilated and the entropy of the system must be 0. In this case, the system cannot always satisfy the second law of thermodynamics. To avoid the violation of the second law of thermodynamics, we assume that there is a critical particle number density $n_c$ for dust, at which the interaction will be terminated. We can call it the truncation of the interaction between dust and the space-time background. After the particle number density of dust reaches the critical value, they will expand freely. From the previous research, we have known that the entropy of dust always increases if it expands freely. Therefore, as long as the entropy of dust keeps increasing before reaching the critical condition, the interaction between dust and the space-time background will not violate the second law of thermodynamics. Next, we analyze the entropy evolution of dust in three cases.

{\bf Case I:} If the annihilation rate of dust particles is much slower than the expansion rate of the universe ($0<-\frac{N'}{N}\lll \frac{V'}{V}$), in order to determine the sign of Eq.~(\ref{211123www}), we need to estimate the value of $\ln\left[\frac{V}{N}\left(\frac{2\pi k\, m_0 T_0}{h_0^2}\right)^{3/2}\right]$. Let $m_0$ be the average mass of a baryon, which is approximated as $m_0\sim1.6\times10^{-27}$ kg. We assume that the temperature of dust in the universe is $\sim 1$ K. It can be found in the following discussion that the temperature of dust has little effect on the final result. If the particle number density of dust is equal to the particle number density of baryons (which is about $0.2$/m$^3$ at present), then we can take $\frac{V}{N}\sim 5$\,m$^3$. Plugging the Boltzmann constant and Planck constant into $\ln\left[\frac{V}{N}\left(\frac{2\pi k\, m_0 T_0}{h_0^2}\right)^{3/2}\right]$, we finally get
\begin{eqnarray}\label{211123ww2233w}
	\ln\left[\frac{V}{N}\left(\frac{2\pi k\, m_0 T_0}{h_0^2}\right)^{3/2}\right]\sim
	\ln\left[5\left(\frac{2\pi \times 1.6\times10^{-27}\times 1.38\times10^{-23} }{(6.6\times10^{-34})^2}\right)^{3/2}\right]\sim62.
\end{eqnarray}
Note that with the annihilation of dust and the expansion of the universe, $\frac{V}{N}$ will grow and so the value of the above expression will be larger. But due to the logarithm to $\frac{V}{N}$, the increase rate of Eq.~(\ref{211123ww2233w}) is extremely slow. For example, even though $\frac{V}{N}=5\times10^{26}$\,m$^3$, Eq.~(\ref{211123ww2233w}) is only about 122. For the same reason, the temperature of dust has little effect on the value of Eq.~(\ref{211123ww2233w}). When the temperature changes from $10^{-10}$ K to $10^{10}$ K, it only increases from 28 to 97. Therefore, if $0<-\frac{N'}{N}\lll\frac{V'}{V}$, from Eq.~(\ref{211123www}) we can deduce that $\frac{\text{d}S}{\text{d}t}>0$ for the general case (such as $5\,\text{m}^3<\frac{V}{N}<5\times10^{26}\,\text{m}^3$, $m_0\sim1.6\times10^{-27}$ kg, and $10^{-10}\,\text{K}<T_0< 10^{10}\,\text{K}$). It is predictable that for general $m_0$ and $T_0$, only when $\frac{V}{N}$ becomes extremely large, $\frac{\text{d}S}{\text{d}t}<0$ could happen.
Therefore, as long as the critical particle number density $n_c$ is not extremely small, the entropy of dust can satisfy the second law of thermodynamics throughout the evolution of the universe.

{\bf Case II:} If the annihilation rate of dust particles is much faster than the expansion rate of the universe ($-\frac{N'}{N}\ggg\frac{V'}{V}>0$), it is not difficult to obtain $\frac{\text{d}S}{\text{d}t}<0$ for the general case. Only when $\ln\left[\frac{V}{N}\left(\frac{2\pi k\, m_0 T_0}{h_0^2}\right)^{3/2}\right]<-\frac{3}{2}$, the entropy of dust can satisfy $\frac{\text{d}S}{\text{d}t}>0$. Since the entropy of dust is always positive definite:
\begin{eqnarray}\label{46746813}
	S= N\,k\ln\left[\frac{V}{N}\left(\frac{2\pi k\, m_0 T_0}{h_0^2}\right)^{3/2 }\right]+\frac{5}{2}N\,k>0,
\end{eqnarray}
we finally obtain that $\frac{\text{d}S}{\text{d}t}>0$ requires
\begin{eqnarray}\label{6546546}
	-\frac{5}{2}<\ln\left[\frac{V}{N}\left(\frac{2\pi k\, m_0 T_0}{h_0 ^2}\right)^{3/2}\right]<-\frac{3}{2}.
\end{eqnarray}
In the subsequent study, we will see that this condition is unlikely to happen in the realistic universe. Therefore, in this case the entropy of dust usually violates the second law of thermodynamics.

\begin{figure}
	\centering
	\includegraphics[width=15cm,height=10cm]{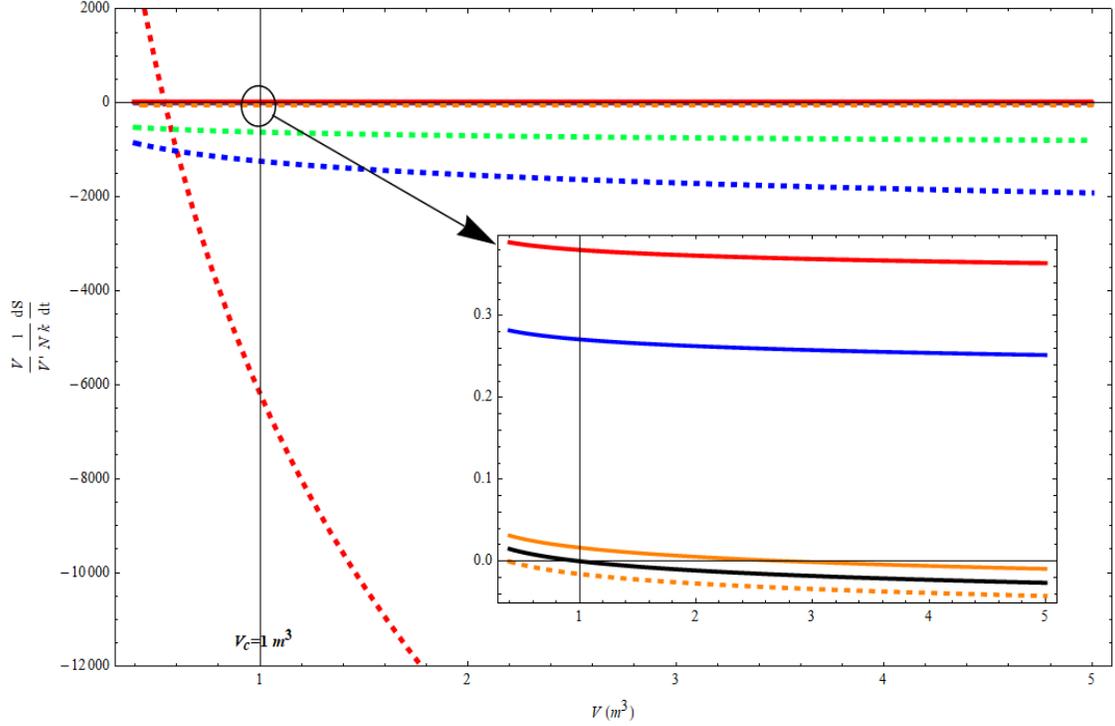}\\
	\caption{Plot of the entropy increase rate of dust with the co-moving volume in the case of dust annihilation. There are eight values of the parameter $b$: $b=0.01$ (red dashed line), $b=0.05$ (blue dashed line), $b=0.1$ (green dashed line), $b=61$ (orange dashed line), $b=61.955$ (black solid line), $b=63$ (orange solid line), $b=85$ (blue solid line), and $b=100$ (red solid line). The entropy increase rates of all dashed lines are negative at $V_c=1\,\text{m}^3$. The entropy increase rate of the black solid line ($b=61.955$) is exactly 0 at $V_c=1\,\text{m}^3$. The entropy increase rates of all other solid lines are positive at $V_c=1\,\text{m}^3$.}\label{p22}
\end{figure}

{\bf Case III:} If the annihilation rate of dust particles is not much different from the expansion rate of the universe, we can set $\frac{V'}{V}=-b\frac{N'}{N}$. Then Eq.~(\ref{211123www}) can be rewritten as
\begin{eqnarray}\label{211123www324}
	\frac{V}{V'}\frac{1}{N\,k}\frac{\text{d}S}{\text{d}t}=\left\{
	-\frac{1}{b}\ln\left[V^{1+\frac{1}{b}}\left(\frac{2\pi k\, m_0 T_0}{h_0^2}\right)^{3/2}\right]-\frac{3}{2b}+1\right\},
\end{eqnarray}
where we have $\frac{V}{V'}\frac{1}{N\,k}>0$. For convenience, we assume that the range of the parameter $b$ is $0.01 \leq b\leq100$, and we choose some specific values of $b$ to examine the entropy evolution of dust. As shown in Fig.~\ref{p22}, for any value of $b$, the entropy increase rate of dust decreases monotonically with the volume $V$. When $V$ is large enough, the entropy increase rate will be negative, which violates the second law of thermodynamics. Moreover, note that when $V$ tends to 0, the entropy increase rate of dust is always positive. Therefore, if the truncation of the interaction occurs early enough, the entropy of dust will not violate the second law of thermodynamics for any value of $b$. We assume that the critical particle number density $n_c$ for dust corresponds to the volume $V_c=1\,\text{m}^3$. Then it can be seen that, only when $61.955<b<100$, the derivative of entropy can be larger than 0 at $V_c=1\,\text{m}^3$. In other cases ($0.01<b<61.955$), the derivative of entropy will be less than 0 at $V_c=1\,\text{m}^3$. Note that the critical value 61.955 is based on
\begin{eqnarray}
	\left(\frac{2\pi k\, m_0 T_0}{h_0^2}\right)^{3/2}=\left[\frac{2\pi \times 1.6\times 10^{-27}\times 1.38\times10^{-23} }{(6.6\times10^{-34})^2}\right]^{3/2}\,\text{m}^{-3}\sim 1.80\times10^{26}\,\text{m}^{-3}.
\end{eqnarray}

With the analysis of the three cases, we can conclude that if the interaction between dust and the space-time background leads the number of dust particles to decrease, then there is an upper limit to the annihilation rate of dust, which is determined by the critical particle number density $n_c$ (or the corresponding $V_c$) and the expansion rate of the universe. Beyond the upper limit, the entropy evolution of dust will violate the second law of thermodynamics.

When $\bar E>0$ (dust absorbs energy from the space-time background), the total number of dust particles inside the co-moving volume will increase, i.e., $\Delta N>0$. The difference in the specific entropy of dust between the initial and final states is also given by Eq.~(\ref{22223www11111}) but with $\Delta N>0$. To judge whether the specific entropy is increasing, we have to know more details about the particle production rate of dust and the expansion rate of the universe. Similarly, in the light of Eq.~(\ref{22223www11111}), if the particle number density of dust increases, the specific entropy will decrease and vice versa.

The entropy evolution of dust can be analyzed on the basis of the previous discussion. Review Eq.~(\ref{211123www}) and note that $\frac{N'}{N}>0$. According to the positive definiteness of entropy (see Eq.~(\ref{46746813})) and the first equality in Eq.~(\ref{211123www}), if $\frac{V'}{V}\geq\frac{N'}{N}>0$, then $\frac{\text{d}S}{\text{d}t}>0$ is always true.

\begin{figure}
	\centering
	\includegraphics[width=15cm,height=10cm]{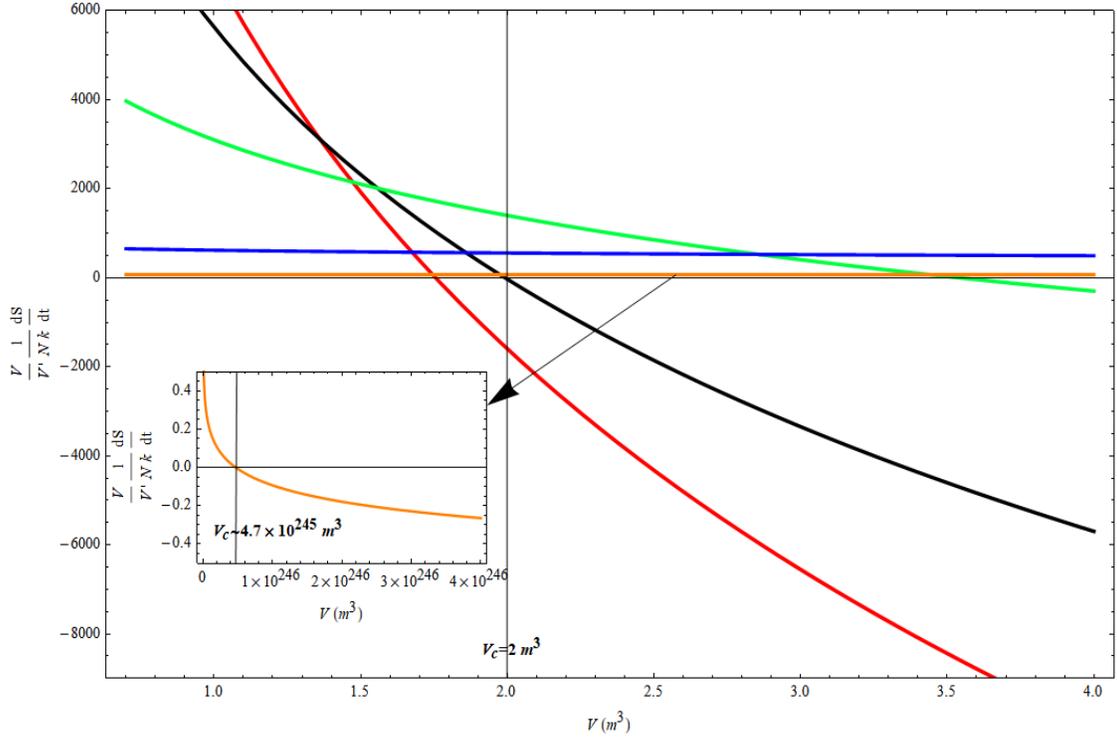}\\
	\caption{Plot of the entropy increase rate of dust with the co-moving volume in the case of dust production. There are five values of the parameter $b$: $b=0.009$ (red line), $b=0.011$ (black line), $b=0.02$ (green line), $b=0.1$ (blue line), and $b=0.9$ (orange line). The entropy increase rate of red line is negative at $V_c=2\,\text{m}^3$. The entropy increase rate of the black line ($b=0.011$) is exactly 0 at $V_c=2\,\text{m}^3$. The entropy increase rates of all other lines are positive at $V_c=2\,\text{m}^3$.}\label{p33}
\end{figure}

If $0<\frac{V'}{V}<\frac{N'}{N} $, we can set $\frac{V'}{V}=b\frac{N'}{N}$ ($V\sim N^b$) with $0< b< 1$. Since $\frac{V}{N}=N^{b-1}$ ($0<b<1$) decreases monotonically with the particle number $N$, with the help of the second equality in Eq.~(\ref{211123www}), it is predictable that only when $\ln\left[\frac{V}{N}\left(\frac{2\pi k\, m_0 T_0}{h_0^2}\right)^{3/2}\right]<0$, we may have $\frac{\text{d}S}{\text{d}t}<0$. However, as stated earlier, this condition is uncommon in the realistic universe, so in this case the entropy of dust usually also satisfies the second law of thermodynamics. Let us briefly analyze why this condition and Eq.~(\ref{6546546}) are uncommon. We rewrite Eq.~(\ref{211123www}) as
\begin{eqnarray}\label{211123www324uy}
	\frac{V}{V'}\frac{1}{N\,k}\frac{\text{d}S}{\text{d}t}=\left\{
	\frac{1}{b}\ln\left[V^{1-\frac{1}{b}}\left(\frac{2\pi k\, m_0 T_0}{h_0^2}\right)^{3/2}\right]+\frac{3}{2b}+1\right\}.
\end{eqnarray}
For any value of $b$ ($0< b< 1$), the entropy of dust will decrease at some point as $V$ grows. Due to the existence of the truncation, if the entropy of dust keeps increasing before the interaction is truncated, then the second law of thermodynamics could be guaranteed. It can be seen from Fig.~\ref{p33} that the larger $b$ is, the larger $V_c$ can be. When $b$ tends to 1 (i.e., $\frac{V'}{V}$ tends to $\frac{N'}{N}$), with the help of Eqs.~(\ref{46746813}) and (\ref{211123www}), one can find that $V_c$ can be infinite. For example, when $b=0.9$, $\frac{\text{d}S}{\text{d}t}=0$ corresponds the volume $V\sim 4.7\times10^{245}\,\text{m}^3$, which means the second law of thermodynamics is always satisfied as long as $V_c< 4.7\times10^{245}\,\text{m}^3$. When $b$ is tiny (i.e., $\frac{V'}{V}$ is much less than $\frac{N'}{N}$), to guarantee that the entropy of dust satisfies the second law of thermodynamics, $V_c$ must be small enough. If $V_c\leq1\,\text{m}^3$, for any value of $b$ ($0< b< 1$), the second law of thermodynamics is always satisfied. For a given $V_c>1\,\text{m}^3$, the second law of thermodynamics will give a lower bound on $b$. For example, if $V_c=2\,\text{m}^3$, then $b$ needs to be larger than 0.011 (see Fig.~\ref{p33}). These results do not seem to suggest anything extreme. However, if one calculates the number of dust particles inside the co-moving volume, one can find that the particle number density of dust is unacceptable. According to $\frac{V}{N}=N^{b-1}$, when $b=0.9$ and $V_c= 4.7\times10^{245}\,\text{m}^3$, we have $N=9.3\times10^{272}$. When $b=0.011$ and $V_c=2\,\text{m}^3$, we have $N=2.3\times10^{27}$. In both cases, the particle number densities of dust are unrealistic. Therefore, if $0<\frac{V'}{V}<\frac{N'}{N} $, the entropy of dust under normal circumstances will not violate the second law of thermodynamics.

Based on the above discussion, we can conclude that if the interaction between dust and the space-time background leads the number of dust particles to increase, the entropy of dust in general satisfies the second law of thermodynamics.

\subsubsection{In the absence of particle production}
\label{sec62}

Now we study the entropy change of dust in the absence of particle production. We assume that the initial state of the system is the same as the case of particle production: the rest mass of a dust particle is $m_0$, the temperature of dust is $T_0$, and the co-moving volume is $V_0$. Therefore, the entropy of dust at the initial state is also given by Eq.~(\ref{22223www}). When the energy conversion between dust and the space-time background is equal to $\bar E$, the co-moving volume is still $V_1$. In this case, both the temperature and the rest mass can change over time. For convenience, we can treat these two parameters as a whole. Although the energy of dust is conserved as the universe expands, we cannot give the relationship between $\bar E$ and $m\,T$ (which depends on the specific microscopic process). So, we just denote $m_1\,T_1$ as a function of $\bar E$. Then the entropy of dust at the end of the process can be written as
\begin{eqnarray}\label{22223www22yr33}
	\hat S_1= N_0k\ln\left[\frac{V_1}{N_0}\left(\frac{2\pi k\, m_1 T_1} {h_0^2}\right)^{3/2}\right]+\frac{5}{2}N_0k.
\end{eqnarray}
The entropy change of dust is equal to
\begin{eqnarray}\label{22223www2aa223}
	\hat S_1-S_0= N_0 k\ln\left[\frac{V_1}{V_0}\left(\frac{m_1 T_1}{m_0 T_0}\right)^{3/2}\right].
\end{eqnarray}

When $\bar E<0$, the total energy of dust particles inside the co-moving volume will decrease. Since the energy of dust particles expanding freely is conserved (i.e., $m\, T$ is a constant for dust particles expanding freely), $\bar E<0$ means that $m_1 T_1<m_0 T_0$. Since $m_1 T_1<m_0 T_0$ and $V_1>V_0$, we cannot judge whether Eq.~(\ref{22223www2aa223}) is larger than 0. But if the expansion of the universe and the energy decay of dust particles satisfy $\frac{V_1}{V_0} \left(\frac{m_1 T_1}{m_0 T_0}\right)^{3/2}=1$, the entropy of dust will remain unchanged. If $\frac{V_1}{V_0} \left(\frac{m_1 T_1}{m_0 T_0}\right)^{3/2}>1$, $\hat S_1-S_0>0$ meets the second law of thermodynamics and vice versa. Moreover, the number of dust particles is conserved, so the evolution of the specific entropy is consistent with the behavior of the entropy.

When $\bar E>0$, the total energy of dust particles inside the co-moving volume will increase and we have $m_1 T_1>m_0 T_0$. In this case, since $V_1>V_0$, one can obtain $\frac{V_1}{V_0} \left(\frac{m_1 T_1}{m_0 T_0}\right)^{3/2}>1$ and so $\hat S_1-S_0>0$, which means the second law of thermodynamics is always satisfied. Therefore, the specific entropy of dust will keep increasing.

\subsubsection{Comparison}
\label{sec63}

In this section, we compare the entropy changes of dust obtained in the above two cases, and analyze the difference between these two kinds of entropy changes. Note that we do not know the specific relationship between $\bar E$ and $m\,T$ in the absence of particle production. Therefore, the relationship between $\bar E$ and $m\,T$ is the most critical factor in the comparison of the two kinds of entropy changes.

First, we discuss a special case in which the relationship between $\bar E$ and $m\,T$ could guarantee that the two kinds of entropy changes are equal. Comparing Eq.~(\ref{22223www223}) with Eq.~(\ref{22223www2aa223}), one can find that the equivalence between the two kinds of entropy changes ($S_1-S_0=\hat S_1-S_0$) requires that $m_1 T_1$, $m_0 T_0$, and $\bar E$ satisfy
\begin{eqnarray}\label{22223www2aa2aaa23}
	\frac{\bar E}{m_0\,c^2} \ln\left[\frac{V_1}{N_0+\frac{\bar E}{m_0\,c^2}}\left(\frac{2\pi k\, m_0 T_0}{h_0^2}\right)^{\frac{3}{2}}\right]+\frac{5}{2}\frac{\bar E}{m_0\,c^2}
	= N_0 \ln\left[\frac{N_0+\frac{\bar E}{m_0\,c^2}}{N_0}\left(\frac{m_1 T_1}{m_0 T_0}\right)^{\frac{3}{2}}\right],
\end{eqnarray}
where we have used Eq.~(\ref{22223www11}) to substitute $\Delta N $. Equation (\ref{22223www2aa2aaa23}) indicates that $m_1\,T_1$ is related to the volume ($V_1$) of the system, which contradicts the fact that the energy of dust particles expanding freely is conserved. The contradiction can be comprehended in the following way. For dust with the conserved particle number, if it absorbs the energy $\bar E$, then the temperature and the rest mass should be completely determined. These two parameters are entirely unrelated to the volume of the system, because the free expansion of the system does not change the temperature and so the rest mass. However, $m_1 T_1$ in Eq.~(\ref{22223www2aa2aaa23}) relays on the volume ($V_1$) of the system. Therefore, Eq.~(\ref{22223www2aa2aaa23}) is contradictory with the fact that the energy of dust particles expanding freely is conserved. The contradiction means that the two kinds of entropy changes cannot be consistent.

Next, we qualitatively analyze the difference between these two kinds of entropy changes under different circumstances. Since the initial states of the two cases are the same, we just need to consider the difference in the entropy at the final state between the two cases. Subtracting the entropy corresponding to particle conservation (see Eq.~(\ref{22223www22})) from the entropy corresponding to particle production (see Eq.~(\ref{22223www22yr33})) yields
\begin{eqnarray}\label{aaaasss}
	S_1-\hat S_1=\!\!&&\frac{\bar E}{m_0\,c^2}k\ln\left[\frac{V_1}{N_0+\frac{\bar E}{m_0\,c^2}}\left(\frac{2\pi k\, m_0 T_0}{h_0^2}\right)^{3/2}\right]+\frac{5}{2}\frac{\bar E}{m_0\,c^2}k\nonumber\\
	&&+\,N_0k\ln\left[\frac{N_0}{N_0+\frac{\bar E}{m_0\,c^2}}\left(\frac{ m_0 T_0} {m_1 T_1}\right)^{3/2}\right].
\end{eqnarray}
It can be verified that the above formula is equal to 0 when $\bar E=0$. For a given $\bar E$, since we do not know the specific expression of $m_1 T_1$ with respect to $\bar E$, we can only qualitatively analyze the sign of Eq.~(\ref{aaaasss}). We assume that $\bar E$ is an infinitesimal quantity, that is, the interaction is extremely weak, which is obviously in line with the current observations of the universe. Letting $m_1 T_1$ be an arbitrary function of $\bar E$, when $\bar E=0$, we have $m_1 T_1(\bar E=0)=m_0 T_0$. By expanding $S_1-\hat S_1$ in Eq.~(\ref{aaaasss}) with respect to the parameter $\frac{\bar E}{m_0\,c^2}$ and retaining the first-order term, we have
\begin{eqnarray}\label{aaaaaaawwsss}
	S_1-\hat S_1\sim\frac{\bar E}{m_0\,c^2}k\left\{\ln\left[\frac{V_1}{N_0}\left(\frac{2\pi k\, m_0 T_0}{h_0^2}\right)^{3/2}\right]+\frac{3}{2}-\frac{3}{2}N_0
	\frac{(m_1T_1)'|_{\bar E=0}}{m_0T_0}\right\},
\end{eqnarray}
where $(m_1T_1)'|_{\bar E=0}$ is the derivative of $m_1T_1$ with respect to $\frac{\bar E}{m_0\,c^2}$ evaluated at $\bar E=0$. The value of $(m_1T_1)'|_{\bar E=0}$ reflects the change rate of $m\,T$ with the energy $\bar E$ at the point $m_0T_0$. The key to judging whether the right hand side of Eq.~(\ref{aaaaaaawwsss}) is larger than 0 is to determine $\ln\left[\frac{V_1}{N_0}\left(\frac{2\pi k\, m_0 T_0}{h_0^2}\right)^{3/2}\right]$ and $N_0\frac{(m_1T_1)'|_{\bar E=0}}{m_0T_0}$. It is known from the previous research that $\ln\left[\frac{V_1}{N_0}\left(\frac{2\pi k\, m_0 T_0}{h_0^2}\right)^{3/2}\right]$ is insensitive to $\frac{V_1}{N_0}$ and $m_0 T_0$. For example, even if $\frac{V_1}{N_0}$ takes values from 5\,m$^3$ to $5\times10^{26}$\,m$^3$, the value of $\ln\left[\frac{V_1}{N_0}\left(\frac{2\pi k\, m_0 T_0}{h_0^2}\right)^{3/2}\right]$ (with $m_0\sim1.6\times10^{-27}$ kg and $T_0\sim 1$ K) only increases from 62 to 122. Therefore, whether the right hand side of Eq.~(\ref{aaaaaaawwsss}) is larger than 0 mainly depends on $N_0\frac{(m_1T_1)'|_{\bar E=0}}{m_0T_0}$. For the general case, we still cannot judge the sign of Eq.~(\ref{aaaaaaawwsss}). Nevertheless, we can analyze the following two extreme cases.

When $m\,T$ changes rapidly with the energy $\bar E$ at the point $m_0T_0$ (i.e., $|N_0\frac{(m_1T_1)'|_{\bar E=0}}{m_0T_0}|\gg1$), if $ \bar E>0$, then $N_0\frac{(m_1T_1)'|_{\bar E=0}}{m_0T_0}\gg1$ and so $S_1<\hat S_1$. Moreover, if $\bar E <0$, then $N_0\frac{(m_1T_1)'|_{\bar E=0}}{m_0T_0}\ll-1$ and one can also obtain $S_1<\hat S_1$. Therefore, in this case, $S_1$ is always less than $\hat S_1$.

When $m\,T$ changes slowly with the energy $\bar E$ at the point $m_0T_0$ (i.e., $|N_0\frac{(m_1T_1)'|_{\bar E=0}}{m_0T_0}|\ll1$), if $\bar E>0$, then $0<N_0\frac{(m_1T_1)'|_{\bar E=0}}{m_0T_0}\ll1$ and so $S_1>\hat S_1$. But, if $\bar E<0$, then $0<-N_0\frac{(m_1T_1)'|_{\bar E=0}}{m_0T_0}\ll1$ and we have $S_1<\hat S_1$. Therefore, in this case, the size of $S_1$ and $\hat S_1$ depends on the sign of $\bar E$.

To sum up, when there exits energy conversion between dust and the space-time background, it is unrealistic to keep the two kinds of entropy changes of dust consistent. For $m\,T$ changing rapidly with energy, the entropy change of dust corresponding to particle production is always less than the one corresponding to no particle production whether dust absorbs energy from the space-time background or not. For $m\,T$ changing slowly with energy, when dust absorbs energy, the entropy change of dust corresponding to particle production is larger than the one corresponding to no particle production. However, when the energy of dust is absorbed, the conclusion is opposite. For the general case, we cannot decide which process has the larger entropy change. All discussion in this section applies to the dark matter with similar statistical properties.

\subsection{Dark energy}
\label{sec7}

In this section, we discuss the entropy of dark energy when there exists interaction between dark energy and the space-time background. Since we have assumed that the number of dark energy particles may be a function of temperature, if dark energy interacts with the space-time background, both the particle number and temperature (or rest mass) will change over time and be affected by the interaction. Therefore, there are not two kinds of entropy changes for dark energy. Since it is difficult to calculate quantitatively the entropy change of dark energy in such a complicated situation, we will only qualitatively analyze the entropy change of dark energy in special cases.

Recalling Eq.~(\ref{316aa}), when $m_b T_b=m_a T_a$, it degenerates into Eq.~(\ref{22223www223}) (dust with particle production), and when $\Gamma=0$, it degenerates into Eq.~(\ref{22223www2aa223}) (dust without particle production). Assuming that the energy conversion between dark energy and the space-time background is $\bar E$ during the expansion of the co-moving volume from $V_a$ to $V_b$, the entropy change of dark energy is similar to Eq.~(\ref{316aa}):
\begin{eqnarray}\label{65466}
	\Delta \bar S_{a\rightarrow b}=\!\!&&N_a\exp\left[\int_a^b\bar\Gamma \text{d}t\right] k\ln\left[\frac{V_b}{N_a\exp\left[\int_a^b\bar\Gamma \text{d}t\right]}\left(\frac{2\pi k\, \bar m_b \bar T_b}{h_0^2}\right)^{3/2}\right]+\frac{5}{2}N_a\exp\left[\int_a^b\bar\Gamma \text{d}t\right] k\nonumber\\
	&&-\left(N_a k\ln\left[\frac{V_a}{N_a}\left(\frac{2\pi k\, m_a T_a}{h_0^2}\right)^{3/2}\right]+\frac{5}{2}N_a k\right),
\end{eqnarray}
where $\bar\Gamma$ represents the particle production rate of dark energy due to the expansion of the system and the interaction. The parameter $\bar m_b \bar T_b$ is a function of the energy $\bar E$ and the co-moving volume $V_b$. Note that here the specific entropy of dark matter can still be given as Eq.~(\ref{h91}), but the parameters $N$, $m$, and $T$ are affected by the energy $\bar E$. For dark energy particles expanding freely, we know that under the constraints (\ref{465456465}), the entropy of the system will increase. For dark energy particles interacting with the space-time background, the entropy of the system can also satisfy the second law of thermodynamics under certain conditions.

When $\bar E>0$, the number of dark energy particles at the end of the process satisfies
\begin{eqnarray}\label{46545aaa6465}
	\bar N_b=N_a\exp\left[\int_a^b\bar\Gamma \text{d}t\right]>N_b=N_a\exp\left[\int_a^b\Gamma \text{d}t\right]>N_a,
\end{eqnarray}
where $\Gamma$ is the particles production rate of dark energy in the absence of interaction (see Eq.~(\ref{newparticle})). Moreover, the energy of a single particle should be higher than the case without interaction, i.e., $\bar m_b \bar T_b> m_b T_b$. With Eq.~(\ref{46545aaa6465}), $\bar m_b \bar T_b> m_b T_b$, and the constraints on dark energy particles expanding freely (see Eq.~(\ref{465456465})), we can prove that $\Delta \bar S_{a\rightarrow b}>0$ is generally tenable.

Since $\bar m_b \bar T_b> m_b T_b$, it is obvious that
\begin{eqnarray}\label{6546aaa6}
	\Delta \bar S_{a\rightarrow b}>\!\!&&N_a\exp\left[\int_a^b\bar\Gamma \text{d}t\right] k\ln\left[\frac{V_b}{N_a\exp\left[\int_a^b\bar\Gamma \text{d}t\right]}\left(\frac{2\pi k\, m_b T_b}{h_0^2}\right)^{3/2}\right]+\frac{5}{2}N_a\exp\left[\int_a^b\bar\Gamma \text{d}t\right] k\nonumber\\
	&&-\left(N_a k\ln\left[\frac{V_a}{N_a}\left(\frac{2\pi k\, m_a T_a}{h_0^2}\right)^{3/2}\right]+\frac{5}{2}N_a k\right).
\end{eqnarray}
Then we investigate the properties of the following function:
\begin{eqnarray}
	F(X)= X k\ln\left[\frac{V_b}{X}\left(\frac{2\pi k\, m_b T_b}{h_0^2}\right)^{3/2}\right]+\frac{5}{2}X k,
\end{eqnarray}
where $X=N_a\exp\left[\int_a^b\bar\Gamma \text{d}t\right]$. The derivative of $F(X)$ with respect to $X$ is given as
\begin{eqnarray}
	\frac{\text{d}F(X)}{\text{d}X}=k\ln\left[\frac{V_b}{X}\left(\frac{2\pi k\, m_b T_b}{h_0^2}\right)^{3/2}\right]+\frac{3}{2} k.
\end{eqnarray}
As we analyzed earlier, $\ln\left[\frac{V_b}{X}\left(\frac{2\pi k\, m_b T_b}{h_0^2}\right)^{3/2}\right]$ is generally positive except that the particle number density of dark energy is extremely large. Therefore, $\frac{\text{d}F(X)}{\text{d}X}>0$ and $F(X)$ increases with the variable $X$. With Eq.~(\ref{46545aaa6465}), we can finally obtain
\begin{eqnarray}\label{6aa546aaa6}
	\Delta \bar S_{a\rightarrow b}>\!\!&&N_a\exp\left[\int_a^b\Gamma \text{d}t\right] k\ln\left[\frac{V_b}{N_a\exp\left[\int_a^b\Gamma \text{d}t\right]}\left(\frac{2\pi k\, m_b T_b}{h_0^2}\right)^{3/2}\right]+\frac{5}{2}N_a\exp\left[\int_a^b\Gamma \text{d}t\right] k\nonumber\\
	&&-\left(N_a k\ln\left[\frac{V_a}{N_a}\left(\frac{2\pi k\, m_a T_a}{h_0^2}\right)^{3/2}\right]+\frac{5}{2}N_a k\right)=\Delta S_{a\rightarrow b}>0,
\end{eqnarray}
where $\Delta S_{a\rightarrow b}$ is given by Eq.~(\ref{316aa}).

When $\bar E<0$, the number of dark energy particles at the end of the process satisfies
\begin{eqnarray}\label{aa46545aaa6465}
	\bar N_b=N_a\exp\left[\int_a^b\bar\Gamma \text{d}t\right]<N_b=N_a\exp\left[\int_a^b\Gamma \text{d}t\right].
\end{eqnarray}
where $\Gamma$ is still the particle production rate of dark energy in the absence of interaction (see Eq.~(\ref{newparticle})). The energy of a single particle is lower than the case without interaction, i.e., $\bar m_b \bar T_b< m_b T_b<m_a T_a$. In this case, it is not easy to judge the sign of $\Delta \bar S_{a\rightarrow b}$. According to the previous analysis, if the energy of a single particle changes slowly in the process (i.e., $\bar m_b \bar T_b\sim m_b T_b\sim m_a T_a$) and the number of dark energy particles satisfies $\bar N_b=N_a\exp\left[\int_a^b\bar\Gamma \text{d}t\right]>N_a$ (the increase in the particle number caused by the expansion of the system is more than the decrease in the particle number caused by the energy loss), then the entropy of dark energy can keep increasing.

\section{Conclusions}
\label{sec8}
We reviewed thermodynamics of cosmological particle production (annihilation). Combining classical statistics and cosmology, we studied the entropies of photons, dust (baryonic matter), dark matter, and dark energy in the context of the universe.

When these components of the universe are regarded as the perfect fluids expanding freely, we can calculate their entropies and specific entropies separately. For photons, the entropy and specific entropy inside the co-moving volume are both constants (see Eqs.~(\ref{1545}) and~(\ref{15452})). For dust and dark matter, we treated them as a class of substances with similar thermodynamic and statistical properties. We took dust as an example and supposed that their thermodynamic properties are similar to the perfect gas. Then it was found that both the entropy and specific entropy of dust increase with the expansion of the universe (see Eqs.~(\ref{entropychange}) and~(\ref{entropychange3})). Therefore, the entropy of dust satisfies the second law of thermodynamics. For dark energy, since we do not know whether the number of dark energy particles is conserved, without loss of generality, we assumed that the number of dark energy particles inside the co-moving volume evolves with the expansion of the universe (i.e., there is the phenomenon of particle production even if dark energy expands freely). Supposing that they possess the statistical law of a perfect monatomic gas, we found that if the free expansion of the system satisfies Eq.~(\ref{465456465}), then the entropy of dark energy obeys the second law of thermodynamics.

When these components of the universe interact with the space-time background, their entropies and specific entropies will be influenced by the interaction. For photons, if they absorb (lose) energy, the entropy inside the co-moving volume will increase (decreases) with the expansion of the universe. But the specific entropy of photons is still a constant, which is independent on the interaction. For dust and dark matter, if they absorb (lose) energy, there are two possible extreme outcomes. One is that the particle number inside the co-moving volume increases (decreases) and the energy per particle remains unchanged. The other is that the energy per particle increases (decreases) and the particle number inside the co-moving volume is conserved. We calculated the entropy and specific entropy of dust in the two cases, respectively. By analyzing the relationship between the entropy and the parameters, we obtained the conditions for the entropy satisfying the second law of thermodynamics in different situations. When there exits particle production, the evolution of the specific entropy of dust depends on the change in the particle number density. And when there exits no particle production, the evolution of the specific entropy of dust is consistent with the entropy. By comparing the entropy changes of dust in the two cases, we found that the two kinds of entropy changes cannot be identical. The difference of the two kinds of entropy changes mainly depends on two aspects: the sign of the energy conversion and the change rate of $m\,T$ with the energy conversion. Finally, we analyzed simply the conditions for the entropy of dark energy keeping increasing in different situations when there exists interaction between dark energy and the space-time background.

The entropy of each component of the universe is closely related to the evolution of the universe. Studying the entropy of the universe is of significance for us to understand the fate of the universe. For the realistic universe, both particle production and interaction between different substances are common phenomena. In this paper, when we studied the interaction between matter and the space-time background, we did not consider the microscopic process and the entropy of the space-time background, which is the practice of most literature~\cite{Harko:2014pqa,Harko:2015pma,Nunes:2016tsf,Ramos:2017cot,Yu:2018qzl,SolaPeracaula:2019kfm}. If the entropies of two interacting substances can be taken into account at the same time, obviously more convincing results could be obtained, which is an extended topic that we can study in the future.

\section*{Acknowledgments} \hspace{5mm}
This work was supported by the National Natural Science Foundation of China (Grants No. 11875151, No. 12047501, No. 11873001, and No. 12047564), ``Lanzhou City's scientific research funding subsidy to Lanzhou University'', and the Postdoctoral Science Foundation of Chongqing (Grant No. cstc2021jcyj-bsh0124).

\end{document}